\documentclass[prl,twocolumn,showpacs,floatfix]{revtex4}

\newcommand{\Eq}[1]{Eq.~(\ref{#1})}

\newcommand{\Fig}[1]{Fig.~\ref{#1}}
\newcommand{\Figure}[1]{Figure~\ref{#1}}

\renewcommand{\k}{\mathbf{k}}

\newcommand{\R}{\mathbf{r}}
\newcommand{\Vol}{{\cal V}}
\newcommand{\unitv}[1]{\hat{\mathbf{#1}}}
\newcommand{\expt}[1]{\left< #1 \right>}
\newcommand{\Skip}[1]{}

\usepackage{graphicx}

\begin{document}
\title{Pinning and Depinning of the Bragg Glass in a Point Disordered Model
  Superconductor}

\author{Peter Olsson}

\affiliation{Department of Physics, Ume\aa\ University, 
  901 87 Ume\aa, Sweden}

\date{\today}   

\begin{abstract}
  The three-dimensional frustrated anisotropic XY model with point disorder is
  studied with both Monte Carlo simulations and resistively-shunted-junction
  dynamics to model the dynamics of a type-II superconductor with quenched point
  pinning in a magnetic field and a weak applied current. Both the collective
  pinning and the depinning of the Bragg glass is examined. We find a critical
  current $I_c$ that separates a creep region with unmeasurable low voltage from
  a region with a voltage $V\sim I-I_c$, and also identify the mechanism behind
  this behavior. It is further found to be possible to collapse the data
  obtained at a fixed disorder strength by plotting the voltage versus $TI$,
  where $T$ is the temperature, though the reason for this behavior is unclear.
\end{abstract}


\maketitle

The behavior of elastic structures in the presence of point disorder is a
profound question in condensed matter physics with relevance e.g.\ for charge
density waves and vortex lattices. Due to the competition between the repulsive
interactions, which favor a periodic structure, and both thermal fluctuations
and quenched disorder, with the effect to weaken this order, already the static
problem is a very difficult one. The common picture has for some time been that
the quenched disorder turns the vortex lattice into a Bragg glass, which is a
phase with algebraically decaying correlations\cite{Nattermann:90,
  Giamarchi_LeDoussal:94}, though some recent papers suggest a more complicated
phase diagram\cite{Beidenkopf_AMSZRBT, Li_Rosenstein:03}.

The dynamical properties of a point disordered system in the presence of a
strong driving force is an active field with several open questions and
competing scenarios. In contrast, the behavior at weak fields, which is the
subject of the present Letter, has been less debated, and the usual picture is
that of a sharp depinning transition at zero temperature that is rounded through
thermal activation at non-zero $T$.

In this Letter we present results from dynamical simulations on a
three-dimensional (3D) XY model that give evidence for a different and more
interesting behavior. We find a critical current $I_c$ that separates a creep
region with unmeasurably small voltage from a region with a rectilinear
behavior.  This finding is shown to be related to a memory effect in the system
and is thus manifestly an effect of the system being out of equilibrium.  The
obtained voltage may furthermore be collapsed in an unexpected way. Our
simulations are in many respects similar to Ref.\
\onlinecite{Hernandez_Dominguez}, but our longer simulation times make it
possible to probe the behavior with considerably higher precision.

The Hamiltonian of our 3D XY model is
\begin{equation}
  {\cal H}[\theta_i,\Delta_\mu]=-\sum_{\mathrm{bonds}\,i\mu}
  J_{i\mu}\cos (\theta_i-\theta_{i+\hat\mu}-A_{i\mu} - \Delta_\mu/L_\mu).
  \label{Hamilt}
\end{equation}
Here $\theta_i$ are the phase angles, the $\Delta_\mu$ are twist
variables\cite{Olsson:self-cons} in the three cartesian directions, and the sum
is over all links between nearest neighbors. The vector potential $A_{i\mu}$ is
chosen such that $\nabla\times A_{i\mu} = 2\pi f\unitv{z}$. $f = 1/45$
gives one vortex per 45 plaquettes, which is less dense than in earlier
simulations and is chosen with the aim to reduce the effects of the
discretization by the numerical grid.  The system size is $45\times45\times32$
which gives 45 vortex lines. The anisotropy is chosen as $J_{iz}/J = 1/40$ and
the disorder is introduced as point defects as in Ref.\ \onlinecite{Nonomura_Hu}
through a low density ($=1/180$) of plaquettes in the $x$-$y$ plane with four
weak links.  The weak links have coupling $(1-p)J$ instead of $J$. In this study
of the vortex solid we use disorder strenghts up to $p = 0.34$.  By measuring
the energy of the four links around a vortex both for defect positions and
ordinary plaquettes we find that the pinning energy---the energy needed to
displace a vortex from a defect position to an ordinary plaquette---is
$E_\mathrm{pin} \equiv E_v^{0} - E_v^\mathrm{def} \approx 4.1 p$. 

To check for collective pinning we used standard Metropolis Monte Carlo
simulations with fluctuating twists\cite{Olsson:self-cons}; the driven vortex
solid was studied with RSJ dynamics with fluctuating twist boundary
conditions\cite{Kim_Minnhagen_Olsson}.  The equations of motion are as in
\cite{Chen_Hu} and to integrate them we used a second order Runge-Kutta method
with a dimensionless time step of $\Delta t = 0.1$ and typically (1--4)$\times
10^6$ units of time per run. The twist variable in the $z$ direction is kept
fixed, $\Delta_z=0$.

Since the motion of the vortices is linked to the change of the twist variables,
our main quantities are extracted from the behavior of the twist variables
$\Delta_x$ and $\Delta_y$. In the absence of an applied current $\Delta_\mu$
will perform a random walk and the linear resistance is directly proportional to
the diffusion constant,
\begin{equation}
  R_{l,\mu} = \frac{1}{2T} \frac{1}{t}\langle[\Delta_\mu(t) -
  \Delta_\mu(0)]^2\rangle.
  \label{eq:R}
\end{equation}
A vanishingly small $R$ means that the vortex lattice essentially stays fixed.
An applied current gives a force on the vortices, and their motion means a
change in the twist variables, which is seen as a voltage across the system. The
voltage per link is given by
\begin{displaymath}
  V_\mu = \frac{1}{L_\mu} \frac{d\Delta_\mu}{dt}.
\end{displaymath}
To monitor the order we measure the structure factor
\begin{displaymath}
  S(\k) = \left< \left| \frac{1}{f \Vol}\sum_{\R,z} n_z(\R,z)e^{i\k\cdot\R}
    \right|^2 \right>,
\end{displaymath}
where $\Vol=L_x L_y L_z$, and define $S_1$ as the algebraic average of $S(\k)$
for the six smallest reciprocal lattice vectors, $\k\neq0$.  We also trace out
the vortex lines; the notation $\R_i(z)$ is used for the position in the $x$-$y$
plane of vortex line $i$ at plane $z$.  The mean-squared fluctuation of the
vortex lines $\expt{u^2}$, is also determined. The presented data are for a
single disorder realization, but similar results have been obtained with other
disorder realizations.

\begin{figure}[h]
  \includegraphics[width=8cm]{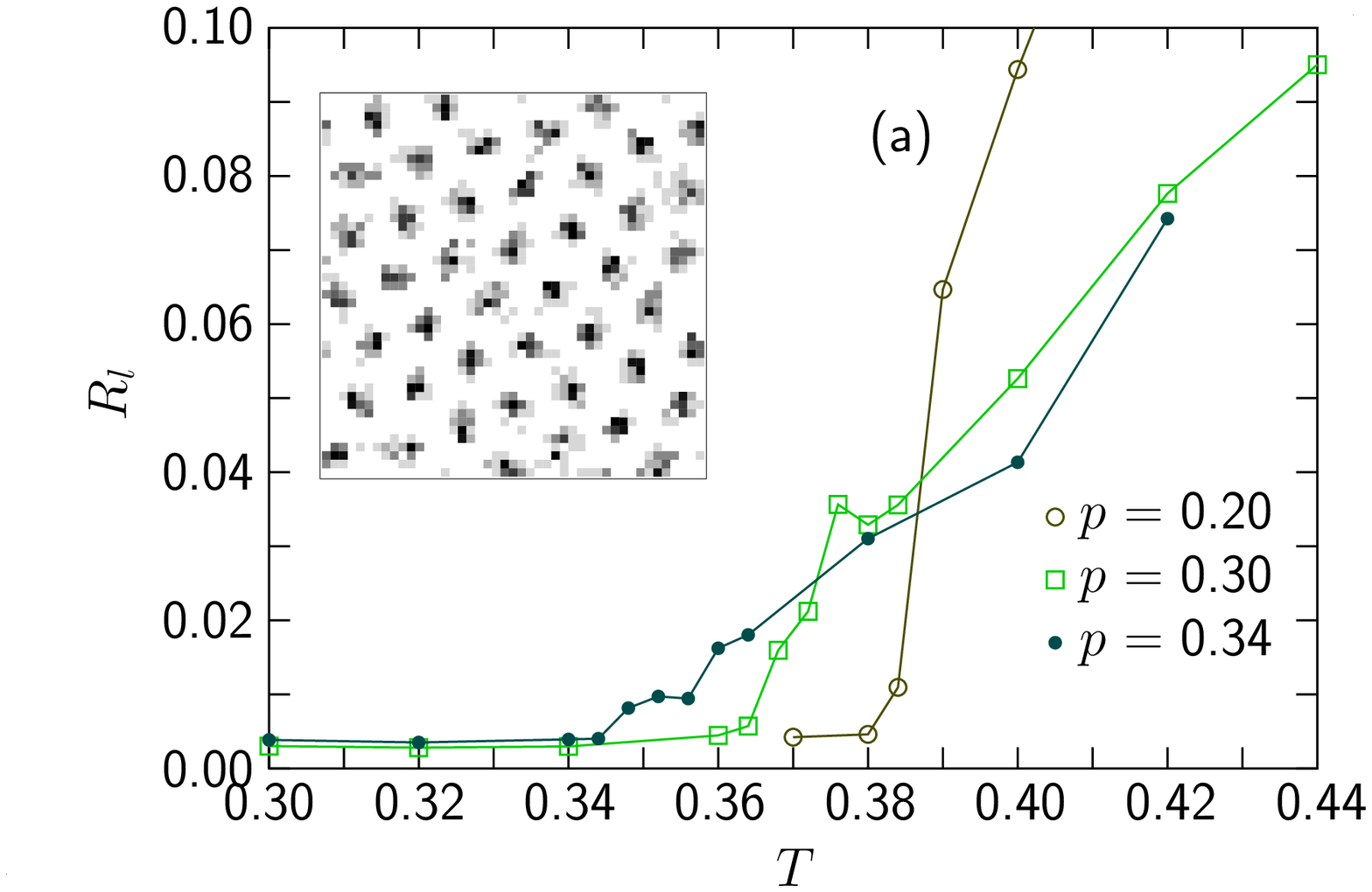}
  \includegraphics[width=8cm]{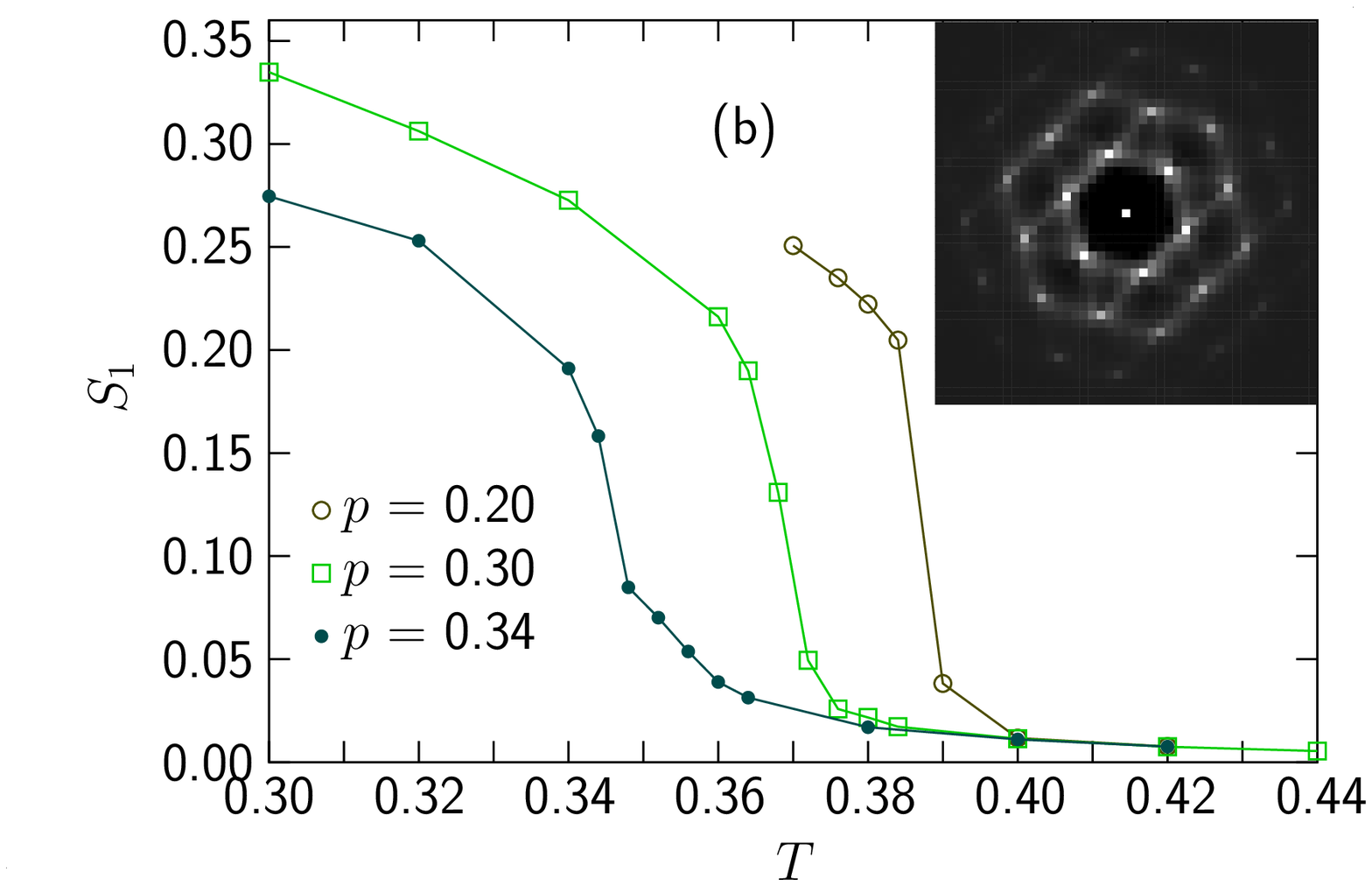}
  \caption{Collective pinning in the vortex solid. Panel (a) shows the linear
    resistance $R_l = \frac{1}{2}(R_{l,x} + R_{l,y})$ determined from \Eq{eq:R}
    with $t = 16384$ from data obtained with MC simulations. (The non-zero $R_l$
    at low $T$ is an artifact of a rather small $t$ together with the
    fluctuations of $\Delta_\mu$ around a fixed values.)  The inset is a top
    view of a configuration of vortices; the darker points indicates a higher
    density.  Panel (b) shows the structure factor $S_1$; the inset is an
    intensity plot of $\ln S(\k)$. Both insets are for $p=0.30$ and $T=0.30$.
    Note that the resistance, and thereby the mobility of the vortex solid,
    decreases rapidly as the temperature drops below melting.}
  \label{fig:xy-dil4R}
\end{figure}

\begin{figure}
  \includegraphics[width=8cm]{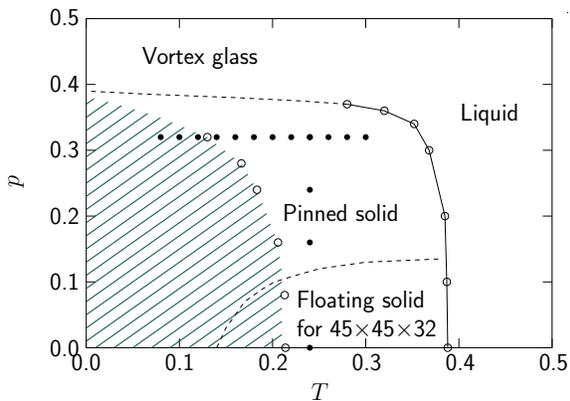}
  \caption{Phase diagram for the simulated model. The open circles connected by
    a solid line is from determinations of melting temperatures from the
    vanishing of the structure factor; the continued dashed line is a sketch of
    a possible behavior at low temperatures. The ``Floating solid'' region at
    low $p$ is where the disorder strength is too weak to pin the vortex solid.
    This is a finite size effect; pinning is restored at a larger size.  The
    filled circles show the location in the phase diagram of the runs in
    \Fig{fig:T240} at $T=0.24$ and the runs in \Fig{fig:pin32} at $p=0.32$. The
    dashed region at low temperatures is where the numerical grid is expected to
    affect the results.}
  \label{fig:PhaseDiag}
\end{figure}

We now turn to the equilibrium dynamics of the point disordered system and show
the temperature-dependence of the linear resistance $R_l$ together with the
structure factor $S_1$ in \Fig{fig:xy-dil4R}. The onset of $R_l$ coincides here
with the vanishing of the structure factor and the direct interpretation is that
the vortex solid is collectively pinned by the disorder whereas the vortex line
liquid is mobile.  The data in \Fig{fig:xy-dil4R} is for rather strong disorder,
$p\geq0.2$. At weaker disorder, $p=0.1$ (not shown), the VL for the size
$45\times45\times32$ is floating whereas collective pinning is restored for a
bigger size, $90\times90\times32$. Based on the data in \Fig{fig:xy-dil4R}
together with some additional simulations at larger $p$, we obtain the phase
diagram in \Fig{fig:PhaseDiag}. The solid symbols are the parameter values where
the dynamic simulations described below have been performed. The dashed region
is where we expect effects of the numerical grid to be important, as discussed
further below.

After having established the existence of collective pinning we turn to our main
subject, which is the dynamics of a weakly driven system. At zero temperature
the vortex lattice depins from the numerical grid at the critical current
density $I_0 \approx 0.13$, but our focus is on depinning from the
\emph{disorder potential} which takes place at current densities more than two
orders of magnitude smaller than $I_0$.  In the simulations the current is
applied in the diagonal direction with $I_x=I_y=I/\sqrt{2}$ since this gives the
force $f_I$ along $-\unitv{v} = (\unitv{x} - \unitv{y})/\sqrt{2}$ which is along one
of the symmetry directions of the vortex lattice, see inset of
\Fig{fig:xy-dil4R}(a). Likewise the voltage per link is $V=(V_x+V_y)/\sqrt{2}$.

\begin{figure}
  \includegraphics[width=8cm]{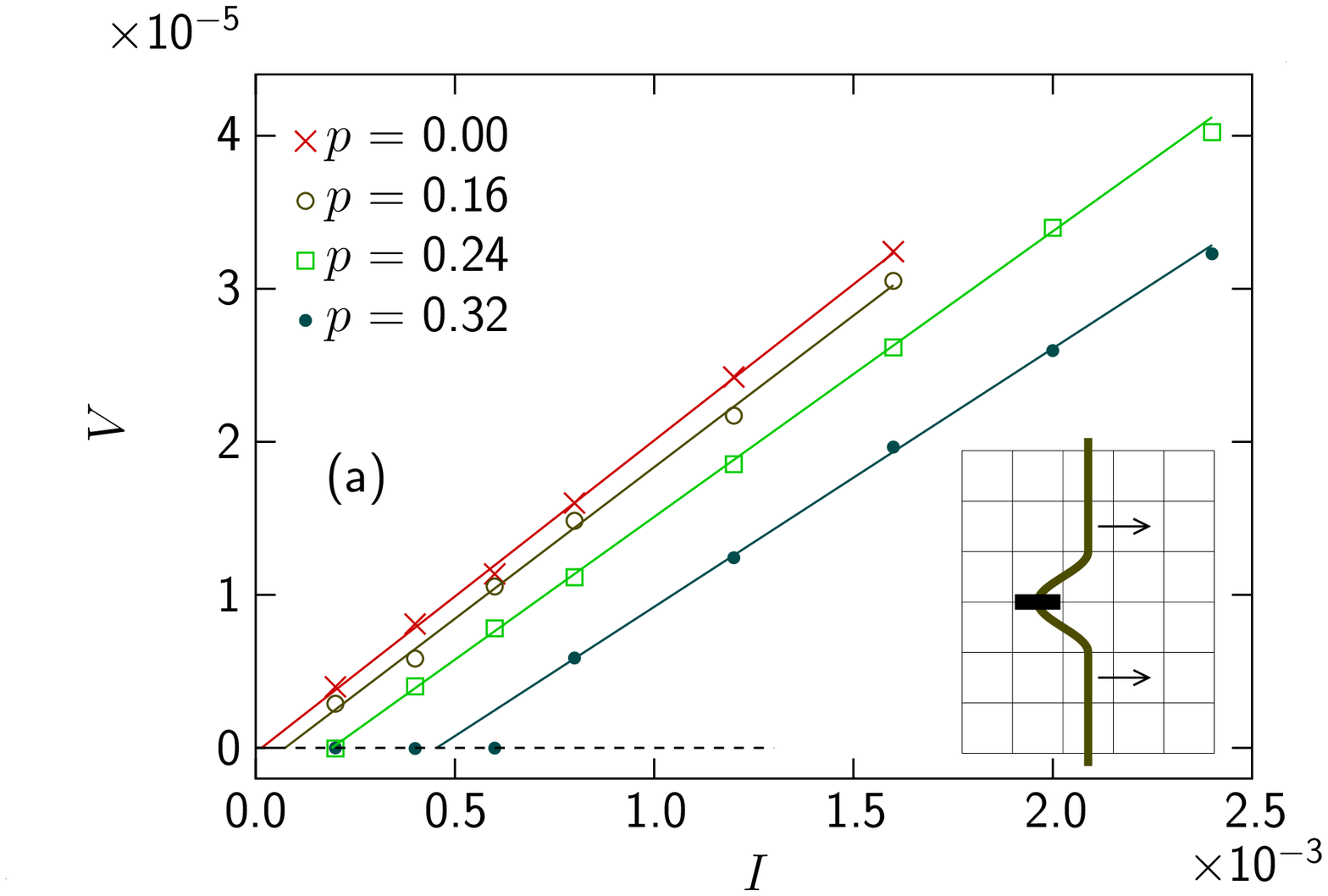}
  \includegraphics[width=8cm]{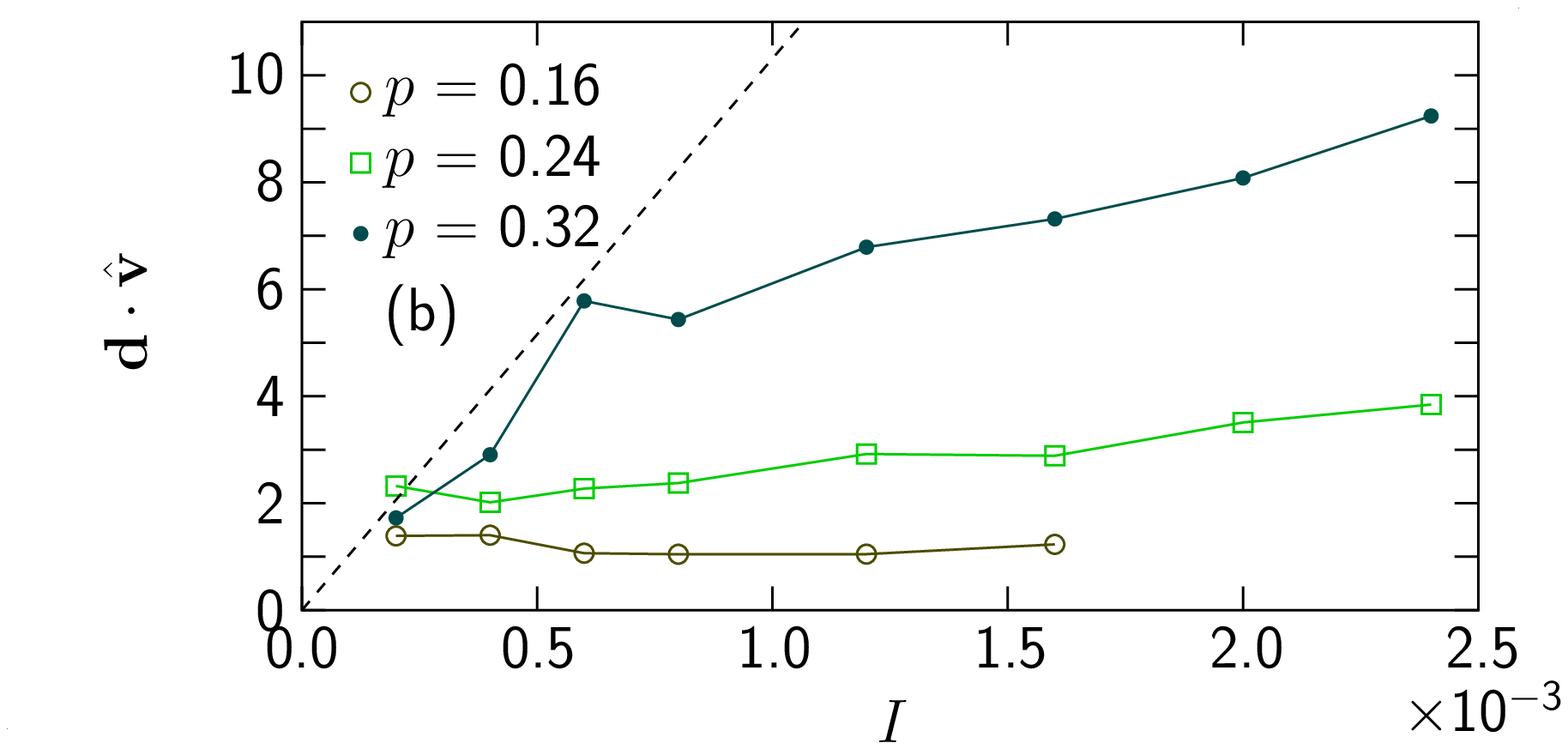}
  \includegraphics[width=8cm]{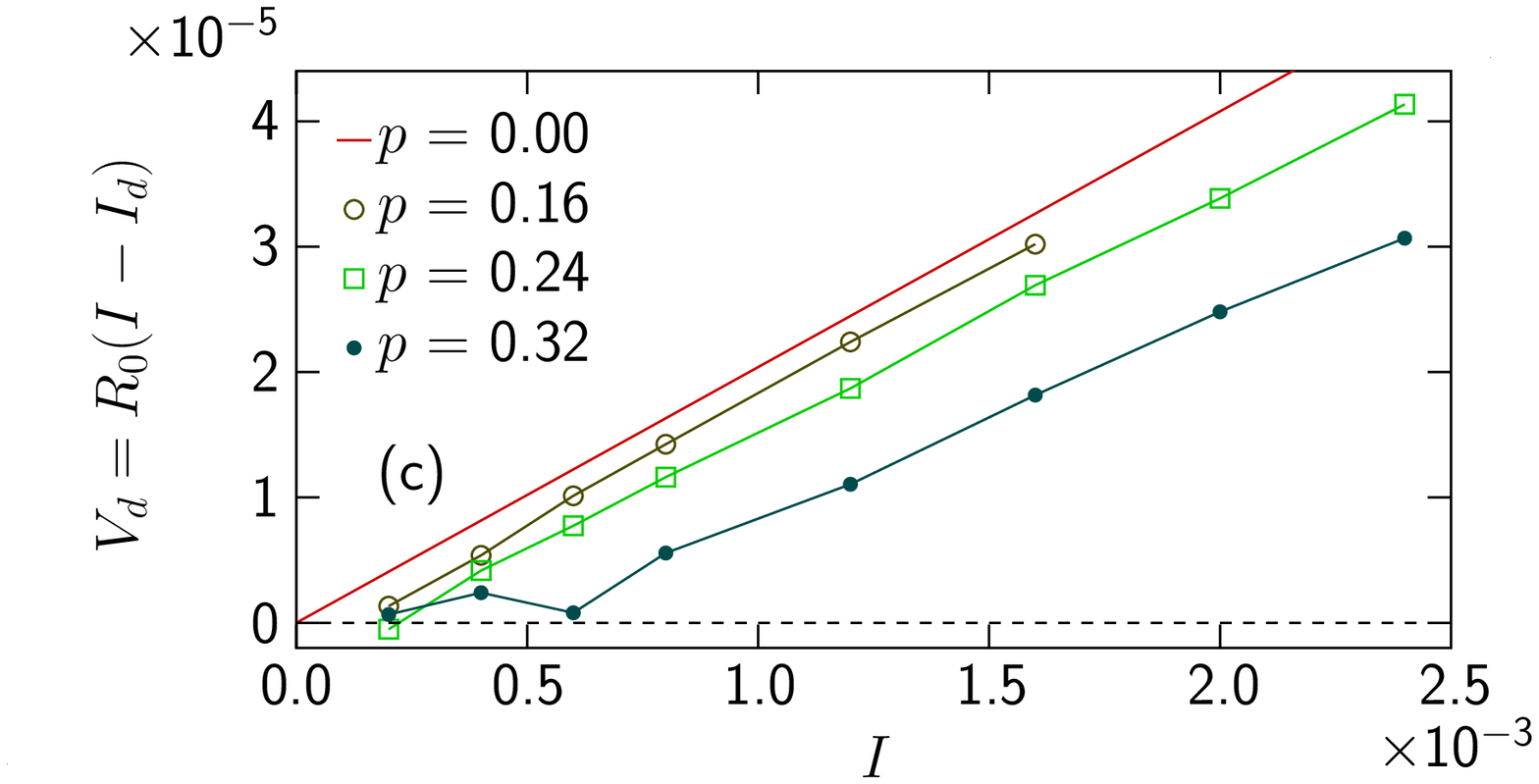}
  \caption{Results at $T=0.24$ and several different $p$. Panel (a) shows that
    the effect of a finite $p$ on the $I$-$V$ characteristics is to shift the
    data down towards zero. This gives a creep region with very low voltage at
    low currents, $I<I_c$, and a region with a rectilinear behavior at larger
    $I$. The inset which is a vortex line extending along $z$ and moving in the
    direction of the applied force, illustrates the suggested mechanism: the
    vortices at defect positions on the average lag after the moving vortex
    lattice.  Panel (b) gives the displacement, c.f.\ \Eq{eq:d} in the direction
    opposite to $f_I$. Panel (c) which is a calculation of the voltage from
    $R_0$ of the pure system and $I_d$ from the measured $d$ and \Eq{eq:Id},
    shows that the suggested mechanism captures the essential elements of the
    dynamics.}
  \label{fig:T240}
\end{figure}

\Figure{fig:T240} shows the $I$-$V$ characteristics at $T=0.24$ with four
different pinning strengths\cite{similarHernandez}. In the pure system, $p=0$,
the voltage is proportional to the current, $V=R_0 I$.  The main effect of
non-zero pinning strength is to shift the voltage down towards zero.  As this
shift downwards only depends weakly on $I$, the slope is almost independent of
$p$.  In this way we get a critical current density $I_c$ that separates the
creep region with an unmeasurably low voltage at $I<I_c$ from the moving region
at $I>I_c$ with $V \sim I-I_c$.

This behavior suggests the existence of a force which, much like a friction
force, has to be overcome to set the vortices in motion and also reduces the
velocity of the moving system. For a system in equilibrium one would expect the
random pinning to contribute with a number of random forces which tend to cancel
one another out, but the slow drift of the vortices opens up for other
possibilities as e.g.\ memory effects. The idea behind our suggested mechanism
is that some of the defect vortices initially stay fixed when the unpinned
vortices slowly move in the direction of the force.  This leads to distortions
of the elastic vortex lines (see inset of \Fig{fig:T240}(a)) and a gradual
build-up of a force $f_\mathrm{el}$ on the defect vortices. This also gives a
reaction force $f_d$ from the defect vortices on the vortex lattice; the motion
of the vortex lattice should then be proportional to $f_I - f_d$.

A basic ingredient in this scenario is that the vortices at the defects on the
average lag after the unpinned vortices. To test this idea we introduce a way to
measure the average displacement of the defect vortices. To distinguish between
ordinary vortices and defect vortices (or rather defect positions) we introduce
the defect operator $D(\R,z)$ which is unity at a defect and zero otherwise. The
average position of all non-defect vortices belonging to the same vortex line
may then be written,
\begin{displaymath}
  \overline{\R}_i = \overline{[1 - D(\R_i(z),z)]\; \R_i(z)},
\end{displaymath}
and as a measure of the lagging behind of the defect vortices, we introduce the
displacement as the sum of their deviations from the average vortex line
positions,
\begin{equation}
  \mathbf{d} = \expt{\sum_{i,z} (\R_i(z) - \overline{\R}_i)\; D(\R_i(z),z)}.
  \label{eq:d}
\end{equation}
Note that this is only the displacements relative to the vortex lines. An
estimate of a total displacement should also include effects due to a vortex
line or a set of vortex lines lagging behind the rest of the vortex lattice.
This is however considerably more difficult to estimate, and as we will see
shortly it seems that the expression above contains the dominant contribution.
Nevertheless, $\mathbf{d}$ is only an approximate estimate---most likely a lower
bound---of the total displacement.

We now expect the displacement vector to be non-zero and point in the direction
opposite to the force. \Figure{fig:T240}(b) which shows how $\mathbf{d}\cdot
\unitv{v}$ depends on the applied current gives strong evidence for the
suggested behavior; the defect vortices on the average lag after the ordinary
vortices.  For a qualitative comparison one would like to compare the magnitude
of this defect force $f_d$ with $f_I$ from the applied current. We do a similar
but more direct comparison by instead estimating $I_d$, the current in the
system due to the displaced vortices. Recall that the creation of a vortex pair
in a 2D model with a separation of $d$ lattice constants in the $y$ direction
creates $d$ rows in the $x$ direction where the phase angle rotates by $2\pi$.
This gives a total current $I_\mathrm{tot} \approx d\sin(2\pi/L_x) \approx 2\pi
d/L_x$.  Taking this over to our 3D system we conclude that a displacement $d$
corresponds to a current density
\begin{equation}
  I_d = \frac{I_\mathrm{tot}}{L_y L_z} \approx \frac{2\pi}{\Vol}d.
  \label{eq:Id}
\end{equation}
At $I<I_c$ we expect $I_d=I$ and it is then possible to obtain the displacement
$d$ from \Eq{eq:Id}. This is the dashed line in \Fig{fig:T240}(b). The data for
$I<I_c$ agrees well with this line. For the behavior at higher currents we plot
in \Fig{fig:T240}(c) $V_d = R_0(I-I_d)$, which is an estimate of the voltage
based on the measured displacement $d$, together with $R_0$ from the pure
system.  The agreement with \Fig{fig:T240}(a) is good and shows convincingly
that our simple description in terms of the displacement of the defect vortices
captures the essential elements of the dynamics.

\begin{figure}
  \includegraphics[width=8cm]{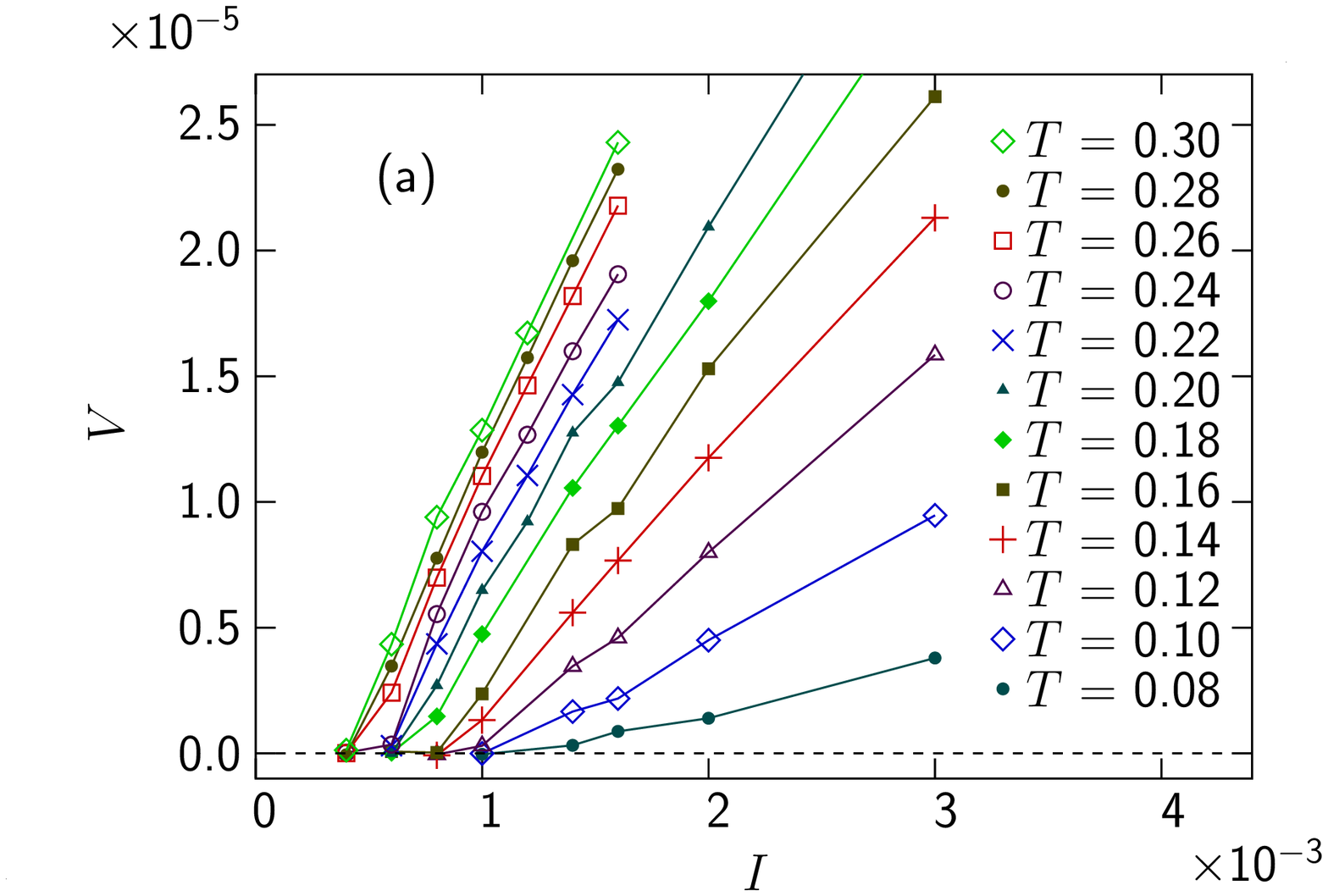}
  \includegraphics[width=8cm]{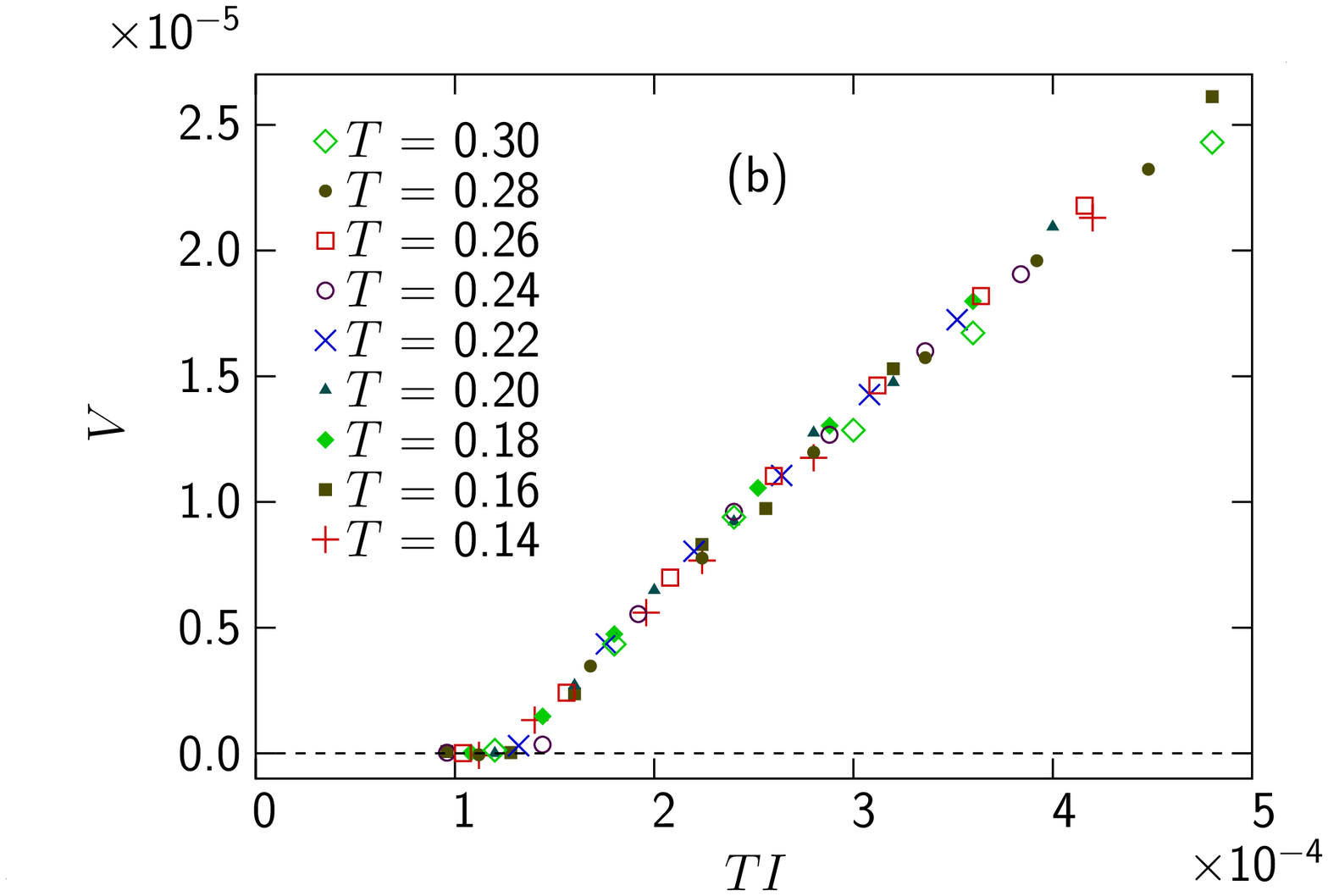}
  \caption{The voltage at $p=0.32$ and several different temperatures.  Panel
    (a) gives evidence for a critical current that increasese with decreasing
    temperature. Panel (b) shows that the voltage may be collapsed to an
    impressing accuracy by plotting the data versus $TI$.}
  \label{fig:pin32}
\end{figure}

We now fix the pinning strength, $p=0.32$, and examine the $I$-$V$
characteristics at several different temperatures from $T=0.08$ to $T=0.30$.
The results are shown in \Fig{fig:pin32}(a) and give evidence for a critical
current that increases with decreasing temperature.  In a further analysis we
have found that it is possible to collapse the data by plotting the voltage
versus the combination $TI$, see \Fig{fig:pin32}(b). This gives $I_c \propto
1/T$ and $V(T,I) = \chi(I/I_c)$ where $\chi$ is a scaling function.  To check if
this behavior is general or depends crucially on our way of introducing the
disorder as a low density of very strong defects, we have done additional
simulations with random couplings as in Ref.\ \onlinecite{Olsson_Teitel:xy3fp}
with $p=0.08$. These simulations give support for the same kind of scaling
collapse.

At low temperatures one expects an effect of the discrete lattice would be to
slow down the dynamics. This is precisely what is seen at $T\leq0.12$ (the data
fall below the common curve) and this data is therefore discarded to produce the
collapse in \Fig{fig:pin32}(b). To study the effect of the discretization we
return to the behavior of the linear resistance in a clean system. At high
temperatures the linear resistance decreases slowly with temperature down to
$T\approx0.22$ where it starts falling rather rapidly down to $R_l=0$ at
$T=0.14$. This shows that the effects of the discrete grid are significant up to
$T\approx0.22$.  Since the discretization is expected to be unimportant if the
typical wandering of the vortex line is much larger than the lattice spacing, we
examine the mean-squared fluctuation which is $\expt{u^2}=1.03$ at $T=0.22$.  It
also turns out that the data at low $T$ that were discarded in
\Fig{fig:pin32}(b) all have $\expt{u^2}<1$ whereas $\expt{u^2}>1$ at the
temperatures included in the figure.  This suggests $\expt{u^2}\gtrsim 1$ as a
criterion for significant effects due to the discretization (the dashed region
in \Fig{fig:PhaseDiag}), and gives support to our claim that the failure to
collapse the data at low $T$ actually is an effect of the discretization of the
underlying grid.

The origin of the collapse in \Fig{fig:pin32}(b) is surprising as several quantities
that could be expected to affect the mobility change considerably across this
temperature interval.  Examples are the fraction of vortices on defect positions
that decreases from 0.078 at $T = 0.14$ to 0.054 at $T=0.30$, the structure
factor that decreases from 0.5 to 0.3, and the resistance in the absence of
disorder which behaves as $R_0(T)\sim T^{1/2}$. 

To conclude, we have obtained the $I$-$V$ characteristics of a point disordered
frustrated 3D XY model and found that the effect of disorder is to give a
critical current that separates the creep region from a region with $V\sim
I-I_c$. The reason for this behavior is a pinning force which appears to be due
to a memory effect; a fraction of the defect vortices lag after the rest of the
vortex lattice. Finally, it is shown that the voltage may be collapsed by
plotting the voltage versus $TI$, though the origin of this behavior as yet not
is fully understood.

We acknowledge discussions with S.~Teitel and P.~Minnhagen, support by the
Swedish Research Council contract No.\ 2002-3975, and access to the resources of
the Swedish High Performance Computing Center North (HPC2N).


\begin{thebibliography}{11}
\bibitem[{sim()}]{similarHernandez}
\bibinfo{note}{If the voltage is plotted on a log scale it becomes clear that
  it is very similar to corresponding data in Fig.~2(a) in Ref.\
  \onlinecite{Hernandez_Dominguez}.}
\end{thebibliography}
\end{document}